\documentclass[superscriptaddress, reprint, aps, prl]{revtex4-1}

\usepackage{amsmath}
\usepackage[utf8]{inputenc}
\usepackage{color}
\usepackage{graphicx}

\usepackage[colorlinks=true]{hyperref}
\hypersetup{linkcolor=blue,urlcolor=blue,citecolor=blue}

\newcommand{\si}{Supplemental Material}

\usepackage{color}

\begin{document}

\title{Quantum coherent multi-electron processes in an atomic scale contact}

\author{Peter-Jan Peters} \affiliation{Institut f\"ur Experimentelle und Angewandte Physik, Christian-Albrechts-Universit\"at zu Kiel, 24098 Kiel, Germany}
\author{Fei Xu} \affiliation{Fachbereich Physik, Universit\"at Konstanz, 78457 Konstanz, Germany}
\author{Kristen Kaasbjerg} \affiliation{Center for Nanostructured Graphene, Department of Micro- and Nanotechnology, Technical University of Denmark,  2800 Kongens Lyngby, Denmark}
\author{Gianluca Rastelli} \affiliation{Fachbereich Physik, Universit\"at Konstanz, 78457 Konstanz, Germany}
\author{Wolfgang Belzig} \affiliation{Fachbereich Physik, Universit\"at Konstanz, 78457 Konstanz, Germany}
\author{Richard Berndt} \email{berndt@physik.uni-kiel.de}  \affiliation{Institut f\"ur Experimentelle und Angewandte Physik, Christian-Albrechts-Universit\"at zu Kiel, 24098 Kiel, Germany}

\begin{abstract}
The light emission from a scanning tunneling microscope operated on a Ag(111) surface at 6~K is analyzed from low conductances  to values approaching the conductance quantum.
Optical spectra recorded at a sample voltages $V$ reveal emission with photon energies $h\nu>2eV.$
A model of electrons interacting coherently via a localized plasmon-polariton mode reproduces the experimental data, in particular the kinks in the spectra at $eV$ and $2eV$ as well as the scaling of the intensity at low and intermediate conductances.
\end{abstract}


\maketitle

A biased nanoscale junction between metal electrodes is a useful environment to study links between quantum transport and electrodynamics.
On one hand, very high current densities may be achieved for electrons at energies beyond the Fermi level.
On the other hand, such junctions support localized plasmon modes that drastically enhance electromagnetic fields \cite{lambe_light_1976, berndt_electromagnetic_1993, ata04, romero_plasmons_2006, savage_revealing_2012, marini, vardi_fano_2016}. 
A consequence of this coincidence is that light affects the conductance of the junction \cite{mol91, vol91, guhr_influence_2007}, another one is the emission of photons, which is driven by the shot noise of the current and corresponds to inelastic tunneling processes \cite{berndt_inelastic_1991, schneider_optical_2010, lemoal,kuhnke}.  
Recently, the latter process has been used to electrically drive optical antennas \cite{bharadwaj_electrical_2011, kern_electrically-driven_2015}.

In addition to the emission at energies $h\nu<eV$ (denoted $1e$ light), where $V$ is the applied bias, higher photon energies $h\nu>eV$ ($2e$ light) have also been observed from metallic and molecular junctions \cite{pec98, down02, hoffmann_two-electron_2003, don04, uem07, schull_electron-plasmon_2009, don10, fuj11, schneider_hot_2013}.
For the observations from single-atom junctions in a scanning tunneling microscope (STM), two models have been proposed.
Xu et al.\ \cite{xu_overbias_2014,xu:16} considered a single, local plasmon-polariton mode modelled by an LC circuit, and attributed the emission to the non-Gaussian quantum noise of the current.
Kaasbjerg and Nitzan calculated the current noise to higher order in the electron--plasmon interaction and found that a plasmon-mediated electron--electron interaction is the source of experimentally observed above-threshold light emission \cite{kaasbjerg_theory_2015}.
The interpretation in terms of coherent processes has been challenged on the basis of experimental data from Au junctions prepared by electromigration and heating of the electron gas has been suggested to lead to the emission at $h\nu>eV$\cite{buret_spontaneous_2015}.
Black-body radiation from heated electrons had previously been invoked in Ref.~\cite{down02}.
\begin{figure}[!b]
\includegraphics[width=0.89\linewidth]{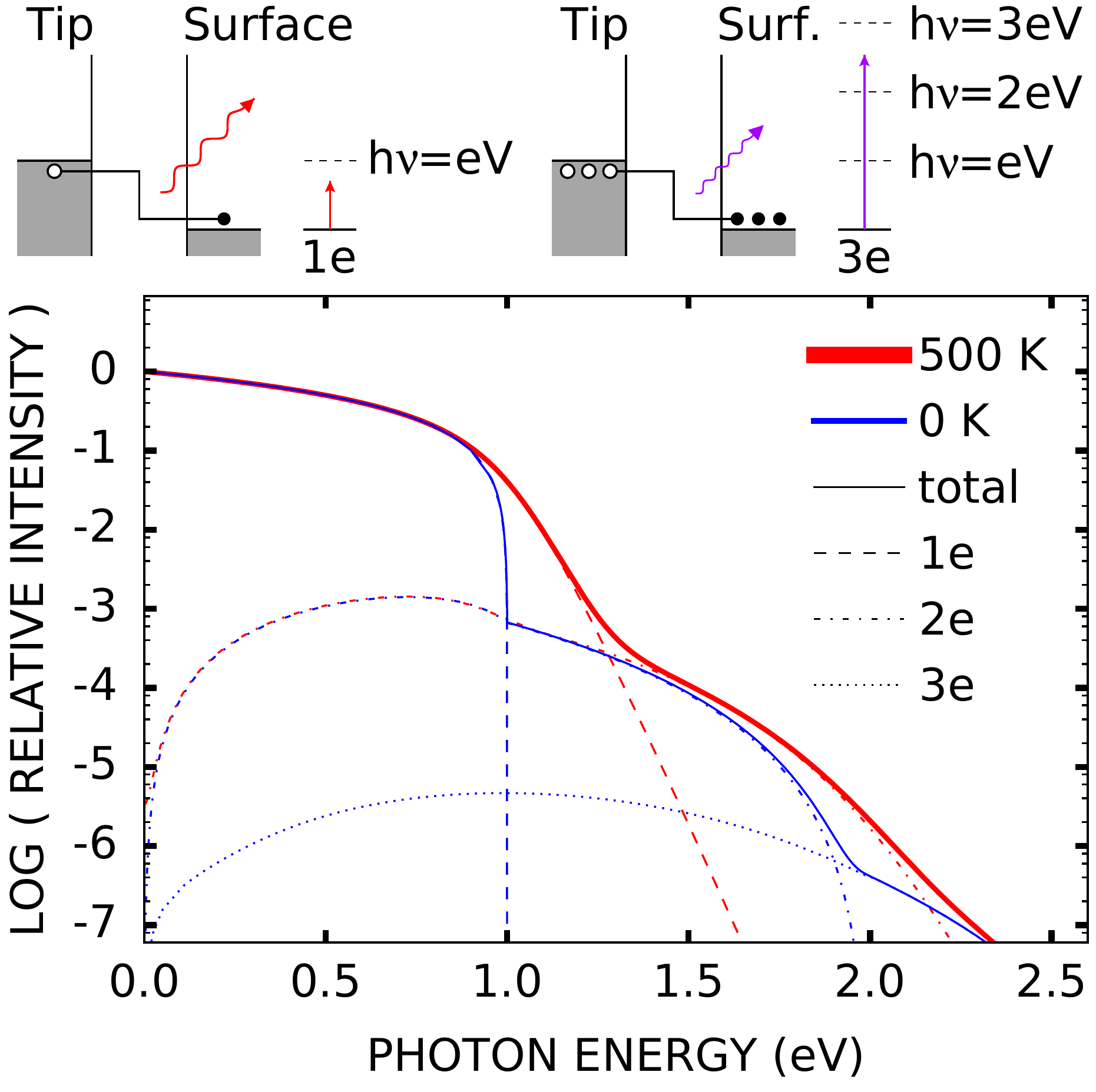}
\caption{Top panel: Energy diagrams for single- and multi-electron photon
    emission processes from a STM junction comprising a tip and a surface.  In a
    conventional $1e$ process (left), tunneling of a single electron leads to
    the emission of a photon whose energy $h\nu$ is limited by the voltage $V$
    applied to the sample.  In a $3e$ process (right), three electrons interact
    via a plasmon to generate a photon with $h\nu$ up to $3eV.$ Bottom panel:
    Calculated emission spectra of multi-electron processes at $T=0$ and 500~K
    and normalized to 1 at zero photon energy. The respective contributions from
    $1e, 2e$, and $3e$ processes are indicated by dashed, dash-dotted, and
    dotted lines. A structureless plasmon spectrum and a coupling parameter
    $\tilde{g}=0.006$ (see text for definition) were used.
\label{overview}}
\end{figure}

In this Letter we report on the emission of $1e, 2e,$ and $3e$ light from
junctions between a STM tip and a Ag(111) single crystal surface.  Optical
spectra reveal kinks at $h\nu=2eV$ and $3eV$ and the emission intensity varies
in a characteristic manner with the junction conductance. The observations
are consistently explained by a model of coherent multi-electron scattering off
the local plasmon field via the higher-order processes sketched in the top
panel of Fig.~\ref{overview}. It leads to the overall behavior of the emission
shown in Fig.~\ref{overview} (bottom) where processes involving $n=1, 2$ and 3
electrons add up to a total intensity, that ---at low temperature--- exhibits
characteristic kinks at the photon energies $h\nu=n eV$, where $V$ is the
sample voltage.  Our work identifies the overbias emission and the
corresponding kinks as distinct fingerprints of higher-order electron-plasmon
scattering processes, and at the same time indicates insignificant electronic
heating in STM contacts.

We used an ultra-high vacuum STM operated at a base temperature of 6~K\@.
Ag(111) samples and electrochemically etched Ag tips were prepared by cycles of Ar ion bombardment and annealing.
After inserting the tips into the STM, they were processed by repeated contact formation at sample voltages $V$ up to 2~V until the tips were stable \cite{castellanos-gomez_highly_2012}. Experiments were conducted on atomically flat terraces.
Light emitted from the tip-sample junction was collected with a lens in-vacuo, then focused onto an optical fiber connected to a grating spectrometer and a thermoelectrically cooled CCD camera \cite{hoffmann_color_2002}. 
The spectrometer/CCD setup could be exchanged for a photomultiplier tube (PMT) for more sensitive measurement. 

\begin{figure}
\includegraphics[width=0.5\textwidth]{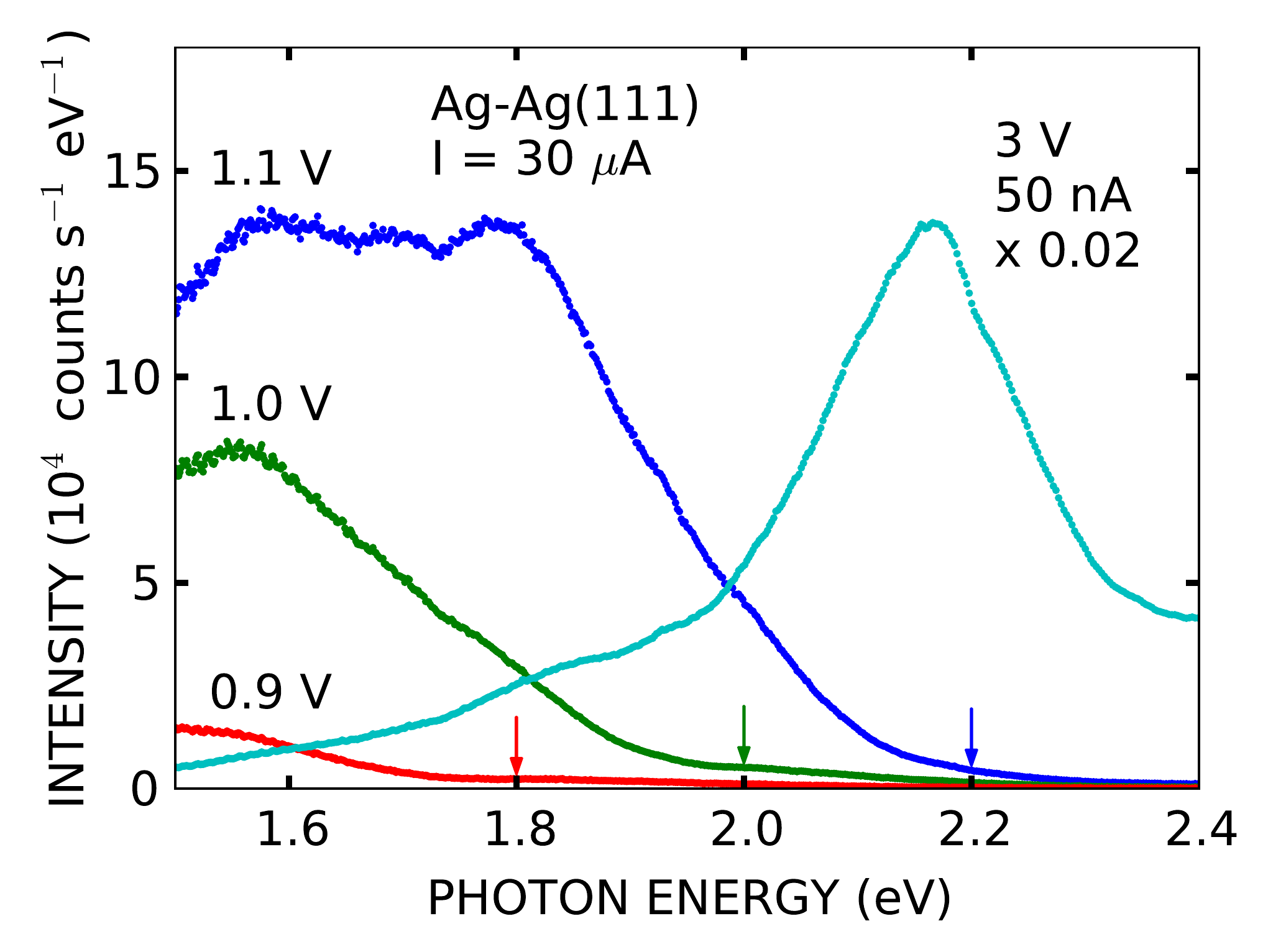}
\caption{
\label{rawspex}
Spectra of the light emitted from a Ag--Ag(111) junction at constant current $I = 30$~$\mu$A for three sample voltages.
A spectrum recorded at 3~V and reduced current is also shown, scaled by a factor of 0.02.
%Integration times: 15 min (2~min for reference spectrum)
Spectra are corrected for dark count rate but not for spectral sensitivity.
Arrows indicate thresholds for $3e$ processes at a photon energy $h\nu = 2 e V$. }
\end{figure}

Figure~\ref{rawspex} shows emission spectra from a Ag--Ag(111) junction recorded at low voltages $V=0.9$--$1.1$~V, %and at 3~V
which were used to keep the photon energy of the more intense $1e$ light below the detection threshold of $\approx 1$eV
\footnote{
The quantum yields of the various emission components depend on the STM tip, the conductance, and the voltage.
We recorded data for $1e, 2e,$ and $3e$ light at $V=3.0, 1.1,$ and 0.8~V, respectively, with the same tip.
Assuming isotropic emission the estimated quantum yield of $1e$ light was $10^{-5}$ at $G=0.02\,G_0$ with this tip.
The yields for $2e$ and $3e$ light at $G\approx 0.4\,G_0$ were lower by factors of $\approx 300$ and $\approx 300^2$, respectively.}.
The spectra reveal significant $2e$ emission, whose intensity drops towards $h\nu=2 eV$ (arrows).
However, light at higher energies $h\nu > 2eV$ ($3e$ light) is also detected.

\begin{figure}
\includegraphics[width=0.89\linewidth]{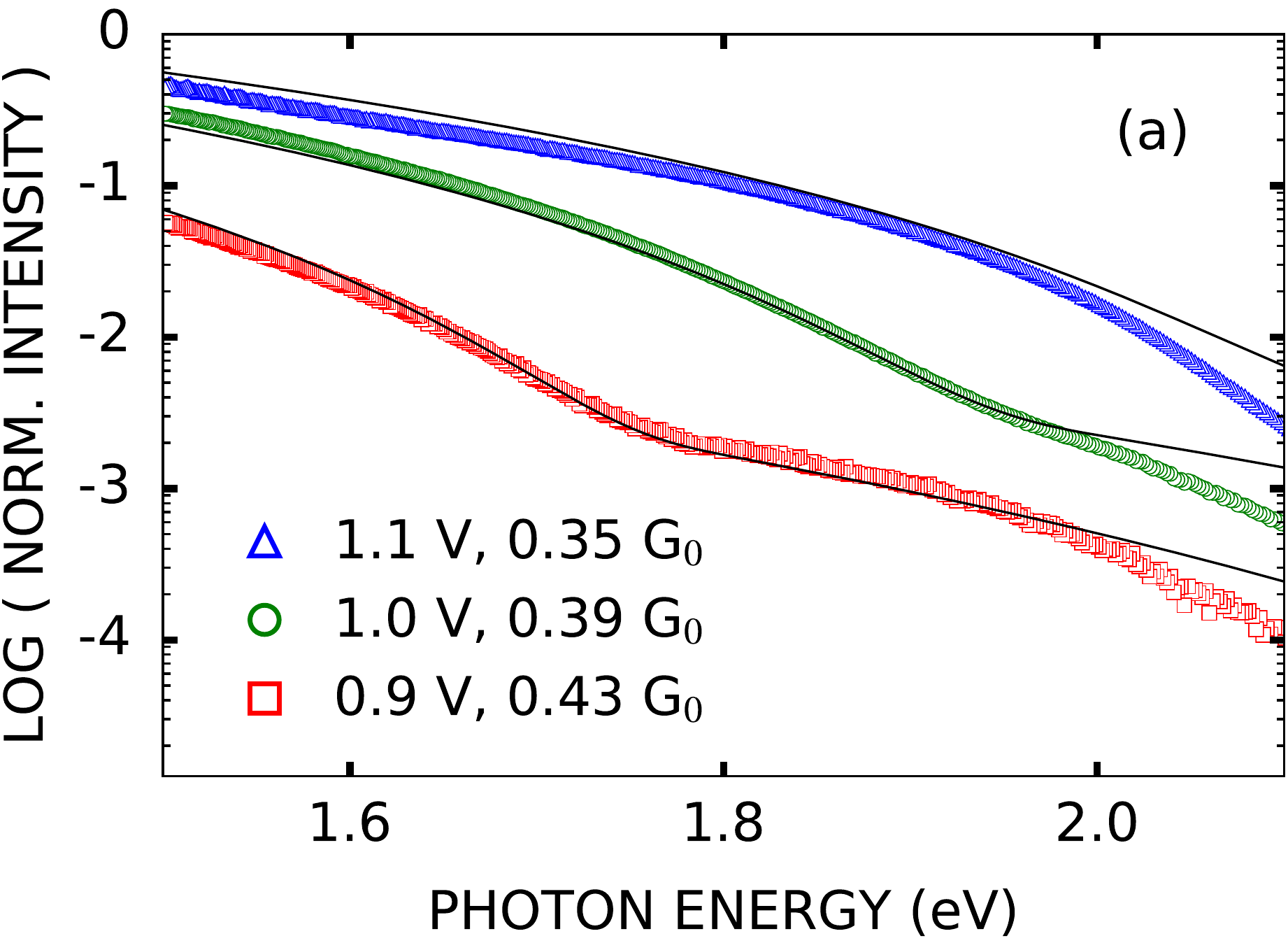}
\includegraphics[width=0.89\linewidth]{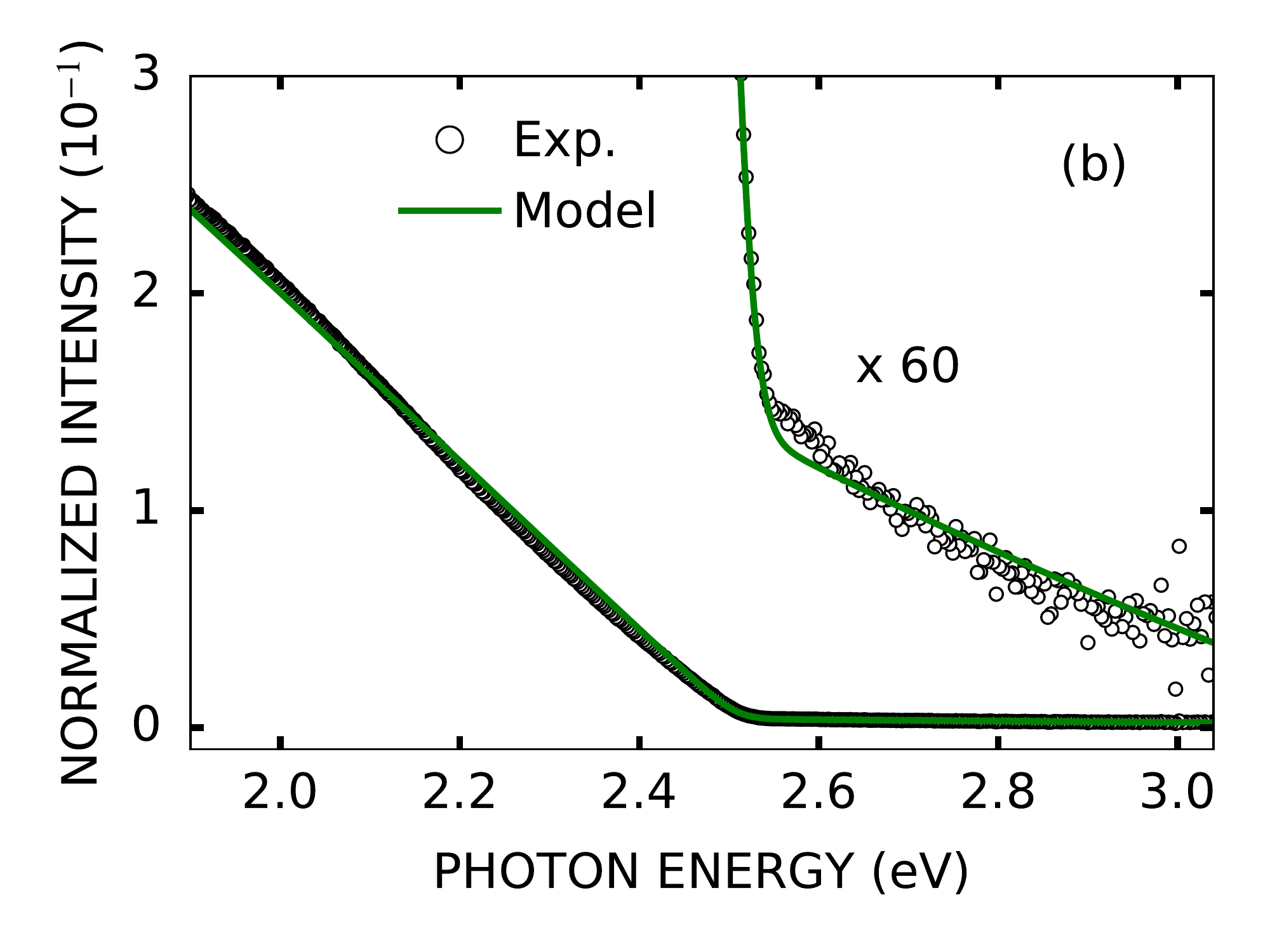}
\caption{
(a) Spectra from Fig.~\ref{rawspex} normalized according to Eq.~\eqref{norm} with the 3~V data and displayed on a logarithmic scale.
Voltages and conductances, in units of $G_0=2e^2/h$, are indicated.
Lines are fits from the model using the experimental conductances and $\hbar\omega_0 = 2.1$~eV, $\eta=0.3$~eV, a temperature of 6~K and coupling parameters $\tilde g=0.013$, 0.03, and 0.05, respectively.
(b) Spectrum of the threshold for $2e$ light measured at 2.5~V normalized with 3.5~V data. 
The spectrum was recorded at $G=0.1\, G_0$ with a tip different from (a).
The line is a fit from the model assuming a temperature of 55~K, which reproduces the position of the kink as well as its rounding. The other parameters are $\hbar\omega_0=2$~eV, $\eta=0.8$~eV, and $\tilde g=0.011$.
Zoomed data has been vertically shifted by 0.1 for clarity. In the whole figure the spectral resolution of the detection setup has
been taken into account.
\label{spex} }
\end{figure}

For further analysis the spectra $R_V$ recorded at a voltage $V$ were normalized by division with a reference spectrum $R_{V_R}$ recorded at an elevated bias $V_R$ while taking into account the expected linear cutoff of $R_{V_R}$ by a weighting factor \cite{lambe_light_1976, aguado_double_2000,blanter_shot_2000}:
\begin {equation}
N_V(h\nu)= \frac{R_V(h\nu)}{R_{V_R}(h\nu)} \left( 1-\frac{h\nu}{eV_R} \right).
\label{norm}
\end {equation}
This procedure removes the effect of the energy-dependent sensitivity of the detection setup \cite{ddeck11}.
It also reduces the influence of the geometry of the junction on a nm scale \cite{rendell_surface_1981, johansson_theory_1990, berndt_inelastic_1991, aizpurua_electromagnetic_2002}.
However, the position and shape of the plasmon resonance to some extent depend on the tip-sample distance \cite{aizpurua_electromagnetic_2002}.
This distance cannot be made identical for measurements at low bias, where $3e$ emission is discernible, and the higher voltages used to determine the shape of the plasmon resonance.
As a result, the normalisation procedure used for Fig.~\ref{spex} does not work well at the onset of the plasmon peak ($h\nu > 2$~eV). 
We verified that the tip shape did not change during a set of measurements by recording reference spectra and STM images before and after.

Figure~\ref{spex}(a) shows normalized data $N_V$.
The intensity, which is mainly due to $2e$ light, smoothly drops towards the threshold for $3e$ light, $h\nu = 2eV$\@. 
Importantly, a clear change of slope is observed near the thresholds, which indicates that an additional radiative process becomes relevant.
A  change of slope is also clearly observed at the transition between the $1e$ and $2e$ spectral ranges shown in Fig.~\ref{spex}(b).
The observed thresholds are difficult to reconcile with the scenario of
Refs.~\cite{down02} and \cite{buret_spontaneous_2015}, where heating of the
electron gas by thousands of Kelvin was invoked to explain the emission of high-energy photons.
That mechanism is not consistent with the present data from STM junctions.

To interpret the light emission up to photon energies of $n e V$, we developed a
model for higher-order scattering processes between $n$ electrons and the dynamic
electric field of a plasmon-polariton resonance by combining the approaches of
Refs.~\onlinecite{xu_overbias_2014, xu:16, kaasbjerg_theory_2015}.  The
tip-induced plasmon of the STM junction is modelled as a damped LC circuit,
which absorbs energy from tunneling electrons and emits photons at energies that
may exceed $e V$.  The electromagnetic enhancement due to the resonance is given
by a Lorentzian $P(E)=\omega_{0}^2/[(\omega_{0}^2-E^2/\hbar^2)^2 + E^2 \eta^2]$, where
$\omega_0=1/\sqrt{LC}$ is the frequency of the plasmon mode and $\eta$ is a
damping parameter. Here $L$ denotes the effective impedance and $C$ the
capacitance.  The resonance describes both the emission enhancement as well as
the effective interaction between the electrons. The coupling coefficient
between the current and the plasmon is expressed as $\tilde g= G G_0L/C$,
where $G$ is the conductance and $G_0=2e^2/h$.

As detailed in the \si, the rate $R_{1e}$ of $1e$ photon emission is governed by
the enhancement factor $P$ and $\tilde S = S/G$ where $S$ is the shot noise
spectral density of the current through non-interacting conductance channels with
transmissions $\{\tau_i\}$ \cite{aguado_double_2000},
\begin{eqnarray}
\label{eq:S}
R(h\nu) & = & R_0 \tilde g P(h\nu) \tilde S(h\nu), \label{eq:R}\\
S(h\nu) & = & G F \left[ w(h\nu-eV)+w(h\nu+eV) \right] \nonumber\\
        & &   + 2 G (1-F) w(h\nu)  ,
\end{eqnarray}
with $G=G_0\sum_i\tau_i$, Fano factor $F=\sum_i\tau_i (1-\tau_i)/\sum_i\tau_i$, and $w(E)=En_B(E)$ where $n_B(E)=(e^{E/k_BT} - 1 )^{-1}$ is the Bose-Einstein distribution. 
$R_0$ is some reference rate, which includes the detector efficiency and other experimentally unkown parameters.

The rate $R_{2e}$ of the $2e$ emission in the frequency range $h\nu > eV$ can at low
temperature, $k_B T \ll eV,h\nu$, be expressed as
\begin{equation}
R_{2e}(h\nu) =  R_0\tilde g^2 P(h\nu)\int\limits_{h\nu-eV}^{eV}d\varepsilon\,P(\varepsilon) 
\tilde S(\varepsilon)\tilde S(h\nu-\varepsilon) ,
\label{eq:Belzig-2e}
\end{equation}
where $\tilde S(\varepsilon)=F(\varepsilon-eV)$ in the integration domain. Correspondingly, the rate $R_{3e}$ of the $3e$ emission in the range $h\nu>2eV$ is
\begin{align}
R_{3e}(h\nu) & =   R_0 \tilde g^3
               P(h\nu)\int\limits_{h\nu-2eV}^{eV}d\varepsilon\int\limits_{h\nu-eV-\varepsilon}^{eV}d\varepsilon' \nonumber\\
  & \quad \times P(\varepsilon) \tilde S(\varepsilon)  
	\tilde S(\varepsilon') P(\varepsilon') \tilde S(h\nu-\varepsilon-\varepsilon').
\label{eq:Belzig-3e}
\end{align}
Full expressions for the rates, in particular at finite temperatures, are provided in the \si.

For illustration, Fig.~\ref{overview} shows the overall spectra calculated assuming a featureless plasmon resonance, viz.\ $P(h \nu)=\mathrm{const.}$
At low temperature, the intensity rapidly drops at photon energies of $neV, \, n=1, 2,$ as expected for processes involving $n$ electrons.
The $1e$ cutoff is less abrupt and shifts to higher energies at an elevated temperature of 500~K because of the broadened Fermi distributions of the electrodes.
Note, that since we only have expressions for the $3e$ light at $T=0$ and in the $3e$ regime, we use always the zero temperature expression and extrapolated the curve beyond the $3e$ regime, but we expect no significant changes in the energy range shown.

Next we use the model to fit experimental spectra.
Figure~\ref{spex}(a) shows a comparison of experimental spectra (symbols) recorded at three voltages and the corresponding calculated results (lines) around the $2e$-$3e$ threshold.
The model reproduces the spectra fairly well assuming $T\approx 20$~K\@.
Deviations mainly occur at $h \nu > 2$~eV, where the normalization of the experimental data is less accurate.

\begin{figure}
\includegraphics[width=0.89\linewidth]{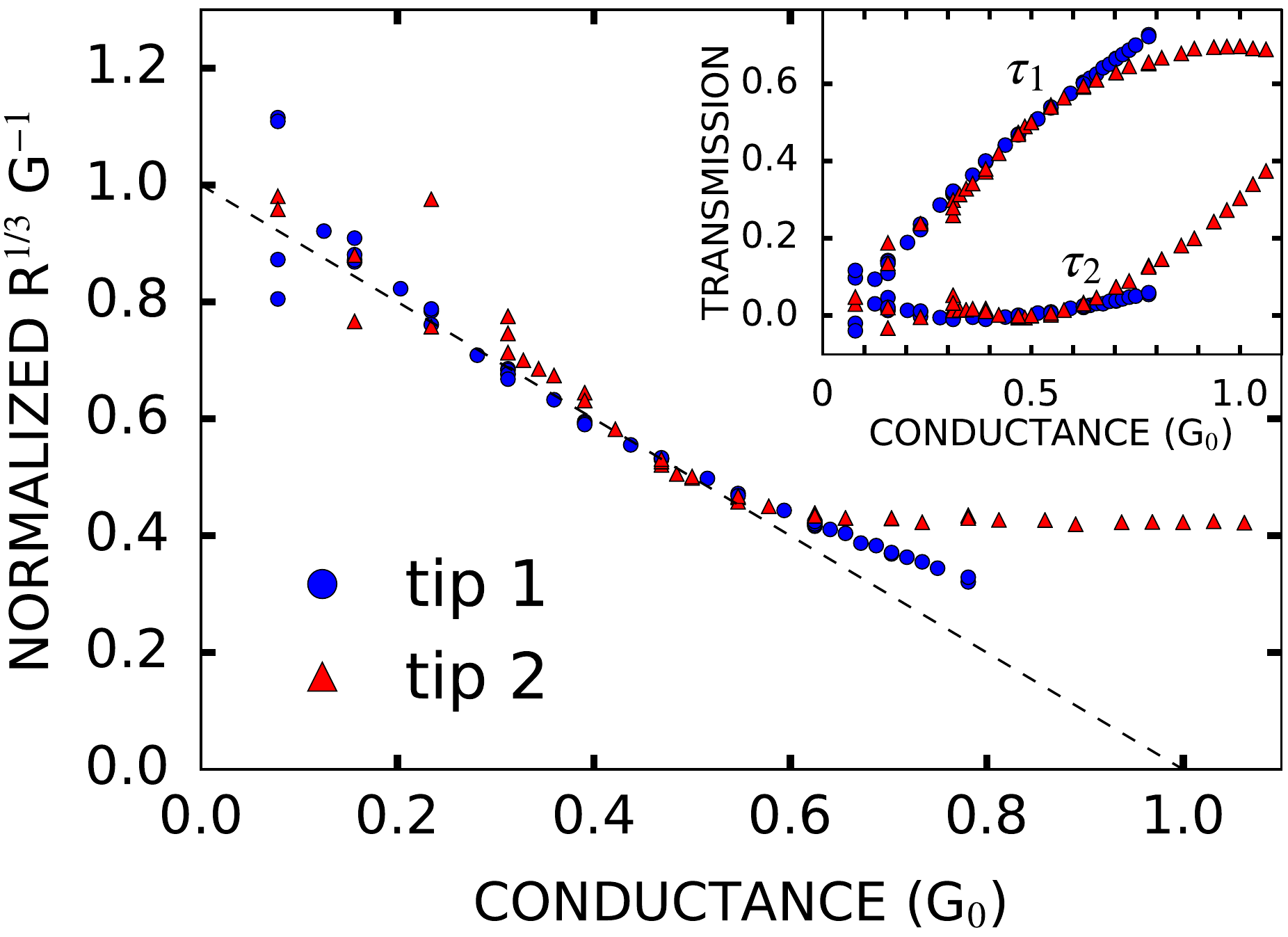}
\caption{Intensity $R$ of light vs. conductance $G$, scaled as $\sqrt[3]{R}/G.$
To reduce the required integration times and thus enable completion of a series of measurements without an unintentional modification of the STM tip the measurements were performed with a PMT\@.  An optical filter was used to limit its sensitivity to photon energies from 1.77 to 3.10~eV\@.  Using a sample voltage $V=0.827$~V only photons with energies $h\nu>2eV$ were detected.
To exclude changes of the tip, approximately the first half of all data were measured at conductances increasing between data points.
The second half was recorded at decreasing conductances to fill in the gaps between data points.
Circles and triangles show data recorded with different tips. 
The dashed line indicates the $F=1-G/G_0$ behavior expected for a
single-channel concact. The inset shows the transmissions for both tips
extracted from a two-channel model with Fano factor~\eqref{eq:twochannel}.
\label{yield}}
\end{figure}

Figure~\ref{overview} suggests that the $1e$-$2e$ threshold should be more well-defined.
This is indeed observed in the data [Fig.~\ref{spex}(b), circles] and reproduced by the model spectra (lines).
In the model, a temperature of 55~K was found to lead to an acceptable fit, demonstrating that some heating does occurs.
A further increase of the temperature would cause a shift of the kink and  additional broadening.
We conclude that the heating is orders of magnitude lower in our STM junctions than that previously invoked to interpret the emission of $2e$ and $3e$ light from electromigrated junctions \cite{buret_spontaneous_2015}.

A fundamental prediction of our theory [cf.\ Eqs.~\eqref{eq:S}--\eqref{eq:Belzig-3e}] is that the low-temperature $ne$ emission in the regime $h\nu>(n-1)eV$ scales with the conductance and Fano factor of a general \emph{multi-channel} contact as
\begin{equation}
  R_{ne} \propto G^n F^n\, .
\label{scaling}
\end{equation}
This suggests that the Fano factor can be inferred from the $ne$ intensity as $F \sim \sqrt[n]{R_{ne}}/G$.
In a simplified scenario of a single conductance channel with transmission
  $\tau$, conductance $G= G_0\tau$, and Fano factor $F= 1-\tau$, a scaling $R_{ne}\propto \tau^n(1-\tau)^n$ previously confirmed for $2e$ emission~\cite{kaasbjerg_theory_2015} is therefore expected.

Figure~\ref{yield} shows the measured photon intensity in the $3e$-light regime $h\nu>2eV$ recorded with two different tips (symbols).
The data have been scaled as $R_{3e}^{1/3}/G$ in order to reflect the Fano factor for a single channel at vanishing conductance.
In addition, the Fano factor for a single-channel contact is shown (dashed line).
Up to conductance of  $\approx 0.5\,G_0$, Fig.~\ref{yield} shows a good match of the data and the simplified single-channel expectation.
However, at higher conductance $G\gtrsim 0.5\,G_0$, significant deviations appear.
To explain the discrepancy, we consider the opening of additional channels with increasing conductance which has previously been observed in Ag and noble metal contacts \cite{scheer_signature_1998,Scheer:2001,lu_light_2013,vardimon_experimental_2013}.
In addition, they presumably are the reason for observed deviations of the
$1e$ yield from a simple $F\sim 1-\tau$ behavior in Ag and Cu contacts \cite{schneider_optical_2010, Abu16}.

The model can be made quantitative by assuming that at most two channels with transmissions $\tau_1$ and $\tau_2$ contribute.
For the conductance and Fano factor we then have
\begin{equation}
  \label{eq:twochannel}
  G = G_0(\tau_1 + \tau_2) \quad \text{and} \quad 
  F = 1 - \frac{\tau_1^2 + \tau_2^2}{\tau_1 + \tau_2}
\end{equation}
respectively.
Considering $\tau_{1,2}$ to be arbitrary functions of the conductance, we can extract their conductance dependencies from the $3e$ intensities in Fig.~\ref{yield} without additional fitting parameters.
The result for the variation of the two channel transmissions with the overall
conductance is shown in the inset of Fig~\ref{yield}. 
The extracted transmission coefficients clearly support the opening of an additional conductance channel at $G\approx 0.5$--$0.6\,G_0$.
The intensity for both tips is thus in full accordance with the model invoking two conductance channels without involving any heating.
Hence, the light emission from the STM can be explained by coherent plasmon-mediated multi-electron processes.

In summary, we have analyzed the emission of $1e$, $2e$, and $3e$ light from atomic scale contacts in a STM\@.
Characteristic features of the spectra, the relative intensities, and the scaling with the conductance are consistently explained in terms of higher-order electron-plasmon interaction.
Our results exclude high electron temperatures as being the reason of the overbias light emission in the present STM experiments.
Rather, they suggest that it is promising to extend the mesoscopic transport
formulation presented here to more complex situations involving, \textit{e.g.},
molecules in the transport path~\cite{schneider_light_2012, reecht_electro_2014, doppagne_vibronic_2017} or complex interacting plasmon resonances.
In antenna structures maximizing the electron-plasmon interaction \cite{parze, bigou} higher-order effects may be further enhanced.
Another interesting direction will be to consider pulsed bias voltages leading to correlated photon emission.

\begin{acknowledgements}
We thank Tom\'a\v{s} Novotn\'y and Juan-Carlos Cuevas for discussions.
K.K. acknowledges support from the European Union's Horizon 2020 research and innovation programme under the Marie Sklodowska-Curie grant
agreement no.~713683 (COFUNDfellowsDTU).
G.R., F.X., and W.B. acknowledge financial support by the DFG through SFB 767.
\end{acknowledgements}

\bibliographystyle{apsrev4-1}

\end{document}